\documentclass[aps,prb,twocolumn,longbibliography]{revtex4-1}

\usepackage{latexsym}
\usepackage[dvipdfmx]{graphicx}
\usepackage{times}
\usepackage{color}
\usepackage{amsmath,amssymb,bm}
\usepackage{multirow}
\begin{document}
\title{Extremely Large Magnetoresistance and Anisotropic Transport in Multipolar Kondo System PrTi$_{2}$Al$_{20}$}
\author{Takachika Isomae$^1$, Akito Sakai$^{2}$,  Mingxuan Fu$^{1,2}$ Takanori Taniguchi$^{3}$, Masashi Takigawa$^{1,4,5}$, and Satoru Nakatsuji$^{1,2,6,7,8}$}
\thanks{Author to whom correspondence should be addressed: satoru@phys.s.u-tokyo.ac.jp}
\affiliation{$^1$Institute for Solid State Physics, University of Tokyo, Kashiwa, Chiba 277-8581, Japan \\
$^2$Department of Physics, Faculty of Science and Graduate School of Science, The University of Tokyo, Hongo, Bunkyo-ku, Tokyo 113-0033, Japan\\
$^3$Institute for Materials Research, Tohoku University, Sendai, Miyagi 980-8577, Japan\\
$^4$Institute of Materials Structure Science, High Energy Accelerator Research Organization (KEK-IMSS), Oho, Tsukuba, Ibaraki 305-0801, Japan\\
$^5$Toyota Physical and Chemical Research Institute, Nagakute, Aichi 480-1192, Japan\\
$^6$CREST, Japan Science and Technology Agency (JST), 4-1-8 Honcho Kawaguchi, Saitama 332-0012, Japan \\
$^7$Institute for Quantum Matter and Department of Physics and Astronomy,
Johns Hopkins University, Baltimore, MD 21218, U.S.A \\
$^8$Trans-scale Quantum Science Institute, University of Tokyo, Bunkyo-ku, Tokyo 113-0033, Japan
}
\date{\today}
\begin{abstract}
\textcolor{black}{Multipolar Kondo systems offer unprecedented opportunities to design astonishing quantum phases and functionalities beyond spin-only descriptions. A model material platform of this kind is the cubic heavy-fermion system Pr$Tr_{2}$Al$_{20}$ ($Tr=$ Ti, V), which hosts a nonmagnetic crystal-electric-field (CEF) ground state and substantial Kondo entanglement of the local quadrupolar and octopolar moments with the conduction electron sea. Here, we explore magnetoresistance (MR) and Hall effect of PrTi$_{2}$Al$_{20}$ that develops ferroquadrupolar (FQ) order below $T_{Q} \sim 2$ K and compare its behavior with that of the non-4$f$ analog, LaTi$_{2}$Al$_{20}$.
In the FQ ordered phase, PrTi$_{2}$Al$_{20}$ displays extremely large magnetoresistance (XMR) of $\sim 10^{3}\%$.
The unsaturated, \textcolor{black}{quasi-linear field ($B$) dependence} of the XMR violates Kohler's scaling and defies description based on carrier compensation alone. By comparing the MR and the Hall effect observed in PrTi$_{2}$Al$_{20}$ and LaTi$_{2}$Al$_{20}$, we conclude that the open-orbit topology on the electron-type Fermi surface (FS) sheet is key for the observed XMR.
The low-temperature MR and the Hall resistivity in PrTi$_{2}$Al$_{20}$ display pronounced anisotropy in the [111] and [001] magnetic fields, which is absent in LaTi$_{2}$Al$_{20}$, suggesting that the transport anisotropy ties in with the anisotropic magnetic-field response of the quadrupolar order parameter. }

\end{abstract}

\maketitle

\textcolor{black}{The quest for material platforms exhibiting large magnetotransport has pushed progress in both fundamental science and technological applications. The outstanding examples, such as giant magnetoresistance in magnetic multilayers, are typically engendered by the interplay of the spin structure with charge transport \cite{Baibich1988,Binasch1989}.
Recent studies uncover XMR in nonmagnetic metals and semimetals, some featuring novel topological band structure, such as Dirac or Weyl nodes\cite{niu2021materials,Takatsu2013,Ali2014,tafti2016resistivity,shekhar2015extremely,narayanan2015linear,leahy2018nonsaturating,yu2017magnetoresistance}.
Nevertheless, a universal understanding of the mechanism behind the observed XMR is lacking. Aside from the spin and charge degrees of freedom, electron orbitals are a critical ingredient for creating new quantum phases and functionalities in strongly correlated systems\cite{tokura2006critical,fernandes2014drives,paschen2021quantum}. Since the electronic band structure finds its root in the interplay between orbitals and the crystal lattice, the ordering and fluctuations of orbitals are expected to yield remarkable effects on transport properties. In 3$d$ transition metal compounds, however, the spin, orbital, and charge degrees of freedom are inextricably intertwined, thus hindering a clear understanding of how orbital ordering and fluctuations tie in with novel transport phenomena. In contrast, cubic 4$f$ rare-earth materials may host a nonmagnetic crystal-electric-field (CEF) ground state with high-rank multipolar moments, offering a route to materialize novel transport phenomena of a purely orbital origin\cite{onimaru2016exotic,patri2020emergent,patri2020critical,sim2020multipolar}.}



The multipolar Kondo system PrTi$_{2}$Al$_{20}$ provides a suitable stage for investigating orbital ordering and its ties to exotic electronic transport. In this system, the cubic $T_d$ symmetry of the Pr site stabilizes a non-Kramers $\Gamma_3$ doublet ground state that carries quadrupolar and octupolar, but no magnetic dipolar moments \cite{Sakai2011}. 
\textcolor{black}{This nonmagnetic ground-state doublet is well separated from the first-excited magnetic triplet by a CEF gap of $\Delta_{\rm CEF}$ $\sim 60$ K,} and thus governs the low-temperature properties of the system \cite{Sakai2011,Kangas2012,Sato2012}. 
A ferroquadrupolar (FQ) order with the order parameter $O_{20}$ ($3J^{2}_{x}-J^{2}$, $3J^{2}_{y}-J^{2}$, and $3J^{2}_{z}-J^{2}$) develops below $T_{Q}\sim 2$ K at zero-magnetic field\cite{Taniguchi2016,Taniguchi2019,Sato2012}, with a superconducting transition inside the FQ phase\cite{Sakai2012}.
Moreover, the cage-like local structure maximizes the number of Al ions surrounding the Pr 4$f$ moments, leading to substantial Kondo entanglement of the multipolar moments with the conduction ($c$) electrons and \textcolor{black}{formation of heavy quasiparticles,} as experimentally confirmed by various experimental probes\cite{Sakai2011,Matsunami2011,Machida2015,tokunaga2013magnetic,dHvA}.
Pressure tuning of PrTi$_{2}$Al$_{20}$ results in a rich phase diagram featuring strongly enhanced superconducting transition temperature $T_c$ and quasiparticle effective mass $m^*$ on approaching the FQ phase boundary and robust non-Fermi-liquid (NFL) behavior over a wide parameter range\cite{Matsubayashi2012}. 
The multipolar Kondo effect and quantum critical fluctuations originating from the orbital degrees of freedom are essential in generating the observed exotic superconductivity and NFL state.

On the other hand, magnetotransport phenomena in PrTi$_{2}$Al$_{20}$ have not been explored.
\textcolor{black}{A giant anisotropic magnetoresistance ratio (AMR) of about $20\%$ is recently reported in the sister compound PrV$_{2}$Al$_{20}$ under a [001] magnetic field\cite{shimura2019giant}, similar to that observed in the nematic order in iron-based superconductors\cite{fernandes2014drives}. This AMR is believed to be driven by quadrupolar (i.e. orbital) rearrangement and the accompanied Fermi surface (FS) change\cite{shimura2019giant}, which opens intriguing prospects of exotic magnetotransport stemming from the interplay of FS properties with the FQ order in PrTi$_{2}$Al$_{20}$.} However, investigations into the FS properties of Pr$Tr_{2}$Al$_{20}$ ($Tr = $ Ti, V) are particularly challenging.  Density functional theory (DFT) calculation of the FS is obscured by the large number of atoms in one unit cell and the strong electronic correlation. Experimentally, a recent de-Haas-van Alphen (dHvA) study on PrTi$_{2}$Al$_{20}$ and LaTi$_2$Al$_{20}$ reveals a complex FS comprising multiple electron and hole sheets in both materials\cite{dHvA}.
\textcolor{black}{
The resolved FS sheets in PrTi$_{2}$Al$_{20}$ shows similar geometry as those in LaTi$_{2}$Al$_{20}$, while
some electron FS sheets in PrTi$_{2}$Al$_{20}$ have enhanced cyclotron effective mass $\sim 8-10m_0$, indicating their sensitivity to the $c$-$f$ hybridization effect\cite{dHvA}.
}
Moreover, NMR\cite{Taniguchi2019}, magnetization\cite{Taniguchi2019}, and specific heat\cite{kittaka2020field} measurements indicate that \textcolor{black}{an applied magnetic field of about 2 T} along certain orientations induces a discontinuous switching of the FQ order parameter, likely accompanied by a change in the $c$-$f$ hybridization which may cause the reconstruction of FS.
\textcolor{black}{
This possible field-induced FS reconstruction due to changes in FQ ordering structure remains an open question; \textcolor{black}{quantum oscillations are observed only for $B$ $\gtrsim$ 2 T, thus offering no evidence for the potential low-field FS changes \cite{dHvA}.}
MR and Hall effect are effective alternatives for probing FS properties; comparing their behavior in PrTi$_2$Al$_{20}$ and LaTi$_2$Al$_{20}$ may yield more profound insights into the interplay of FS and electrical transport properties with multipolar ordering and fluctuations.
}

In this letter, we report transverse MR and Hall effect $\rho_{\rm H}$ in high-quality single-crystal PrTi$_2$Al$_{20}$ and its non-4$f$ analog LaTi$_2$Al$_{20}$. Comparing the observed features for these two materials reveals that open orbits on the electron-type FS sheets are essential for inducing the three orders-of-magnitude increase of transverse MR across the FQ transition in PrTi$_2$Al$_{20}$. Moreover, the MR and Hall effect in PrTi$_2$Al$_{20}$ develop strong anisotropy on approaching the FQ phase, intimately linked to the sharply distinct behavior of quadrupolar order and fluctuations in different magnetic field orientations. \textcolor{black}{The details about material synthesis and experimental methods are in the Supplementary Materials.}

\begin{figure}[t]
 \begin{center}
  \includegraphics[keepaspectratio, width=7cm]{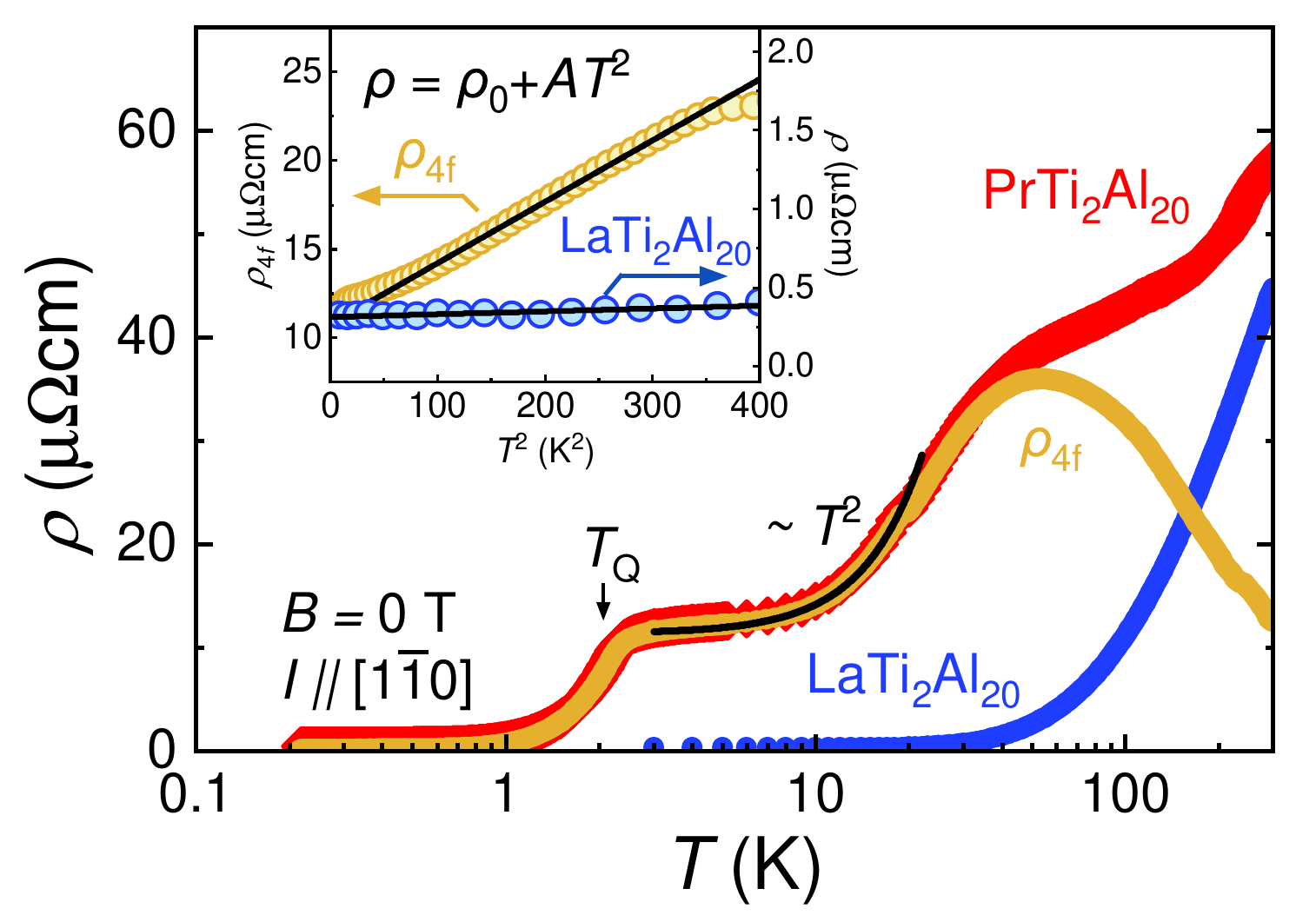}
  \caption{Temperature $T$ dependence of the  zero-field resistivity $\rho$ for PrTi$_2$Al$_{20}$ (solid red circles) and LaTi$_2$Al$_{20}$ (solid blue circles).
  The 4$f$-electron contribution $\rho_{4f}$ (solid yellow circles) is obtained by subtracting $\rho$ of LaTi$_2$Al$_{20}$ from that of PrTi$_2$Al$_{20}$.
  \textcolor{black}{The resistivity of LaTi$_2$Al$_{20}$ is measured down to 2 K.
  To obtain $\rho_{4f}$ below 2 K, we estimated the resistivity of LaTi$_2$Al$_{20}$ below 2 K based on the $T^2$ fit ($\rho$ = $\rho_{0}$ + $AT^2$) shown in the inset, with fitting parameters $A$ = $1.7\times10^4$ $\mu\Omega$cm/K$^2$ and $\rho_{0}$ = 0.31 $\mu\Omega$cm.}
  The downward arrow marks the ferroquadrupolar (FQ) transition temperature $T_Q\sim 2$ K. The solid \textcolor{black}{black} line represents the Fermi liquid behavior $\rho_{4f}\propto T^2$. 
  The inset shows $\rho$ vs. $T^2$ for PrTi$_2$Al$_{20}$ (solid yellow circles) and LaTi$_2$Al$_{20}$ (solid blue circles).
  The solid black lines represent the $T^2$ fits at \textcolor{black}{7 K $\leq T \leq 17$ K for PrTi$_2$Al$_{20}$ and 2 K $\leq T \leq 20$ K for LaTi$_2$Al$_{20}$}.  
  }
  \label{fig_rho}
 \end{center}
\end{figure}

\begin{figure}[b]
\centering 
\includegraphics[keepaspectratio, width = 9cm]{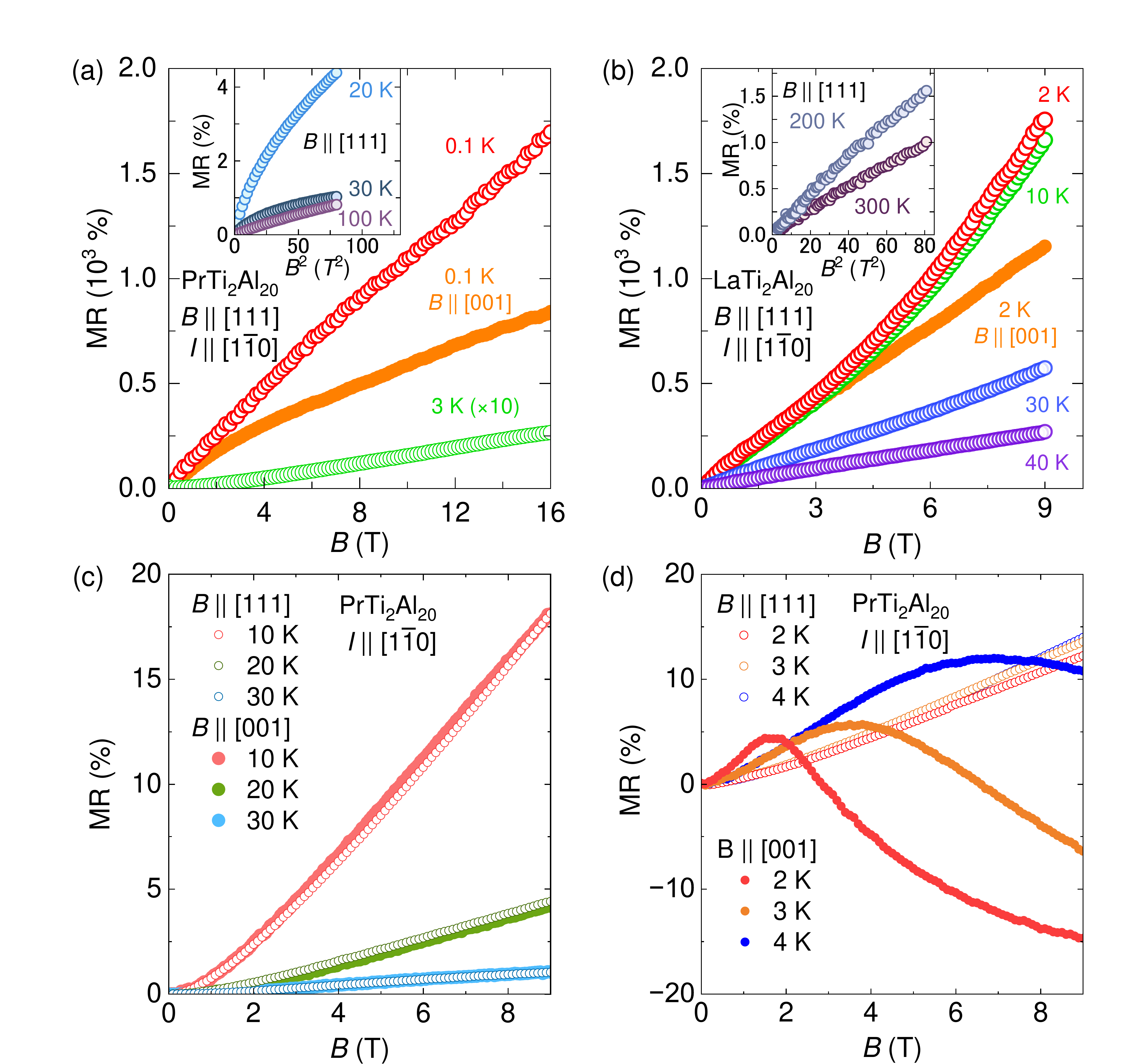}
\caption{
(a) Transverse magnetoresistance (MR) curves of PrTi$_{2}$Al$_{20}$ obtained at representative temperatures.
The main panel shows the MR measured at 0.1 K and 3 K under $B\parallel [111]$ (open symbols) and at 0.1K under $B\parallel [001]$ (orange solid symbols). \textcolor{black}{The 3 K curve is magnified by a factor of ten for clarity.}
The inset shows the MR curves obtained above the FQ transition \textcolor{black}{for $B\parallel [111]$}.
(b) MR curves of LaTi$_{2}$Al$_{20}$ for temperature range from 2 K to 40 K under $B\parallel [111]$ (open symbols) and \textcolor{black}{2 K} under \textcolor{black}{$B\parallel [001]$ (orange solid symbols).
The inset shows the MR curves obtained for $T\gtrsim$ 200 K for \textcolor{black}{for $B\parallel [111]$}.
\textcolor{black}{(c) and (d) Magnetic field $B$ dependence of the MR for PrTi$_{2}$Al$_{20}$ measured in $B\parallel$ [111] (open symbols) and $B \parallel$ [001] (solid symbols) at $T \geq$ 10 K (c) and $T \leq$ 4 K (d).}}
}
\label{fig_MR}


\end{figure}

Figure 1 shows the temperature $T$ dependence of the zero-field resistivity $\rho$ for PrTi$_2$Al$_{20}$ and LaTi$_2$Al$_{20}$.
The zero-field Pr-4$f$ electron contribution to the resistivity, $\rho_{4f}(T)$ is obtained by subtracting the resistivity curve of the isostructural, non-4$f$ analog LaTi$_2$Al$_{20}$ from the raw data of PrTi$_2$Al$_{20}$. \textcolor{black}{In the high-$T$ regime of $T\gtrsim \Delta_{\mathrm{CEF}} \sim 60$ K,} $\rho_{4f}(T)$ shows a logarithmic increase in cooling and reaches a broad peak at $T\sim \Delta_{\mathrm{CEF}}$. The behavior of $\rho_{4f}(T)$ in this $T$-regime is governed by the magnetic Kondo effect arising from the excited magnetic triplets\cite{Sakai2011}.
\textcolor{black}{Below $\Delta_{\mathrm{CEF}}\sim 60$ K, the excited triplet states become less relevant, leading to competing multipolar and magnetic Kondo effects. As a result, $\rho_{4f}(T)$ settles into a Fermi-liquid (FL) regime with $\sim T^{2}$ behavior for 7 K $\lesssim T\lesssim 17$ K,} rather than exhibiting the $\sim {T^{1/2}}$ non-Fermi liquid behavior expected for a quadrupolar Kondo lattice\cite{tsuruta2015non} (Fig. 1, inset).
The $T^2$ coefficient $A$ is about 200 times the value found in LaTi$_2$Al$_{20}$, \textcolor{black}{consistent with heavy fermion formation reported by previous specific heat and dHvA experiments}\cite{dHvA,Sakai2012}.
The sharp exponential decay of $\rho_{4f}(T)$ below $T_Q\sim 2$ K marks the entry into the FQ ordered state with ceased quadrupolar-fluctuation scattering.
\textcolor{black}{We note that $\rho(T)$ deviates from the $T^2$ dependence and shows upward convex curvature $\rho\sim T^n$ ($n \lesssim $ 1) for $T_{\rm Q}<T\lesssim 7$ K.
This behavior can be attributed to a crossover from the FL state driven by competing magnetic and quadrupolar Kondo effects to a non-Fermi-liquid (NFL) state stemming from the quadrupolar Kondo effect, as predicted by numerical renormalization group calculations\cite{kusunose2015competition,kusunose2016competing}.
Previous specific heat measurements reveal an entropy release in the same temperature range, supporting this scenario\cite{Sakai2011}.
In the close neighborhood of $T_{\rm Q}$, critical quadrupolar fluctuations associated with the FQ transition might also influence the behavior of $\rho (T)$, while their effects are unlikely to persist up to as high as 7K.}
The overall behavior of $\rho_{4f}(T)$ is consistent with the previous report\cite{Sakai2011}.

The transverse MR of PrTi$_{2}$Al$_{20}$ measured under $B\parallel[111]$ and $B\parallel [001]$ are shown in \textcolor{black}{Fig. 2(a), (c) and (d)}. 
\textcolor{black}{We first focus on the behavior observed for $B\parallel[111]$.}
In the high-$T$ regime dominated by the magnetic Kondo effect, the MR exhibits quadratic field dependence \textcolor{black}{$\mathrm{MR}\propto B^2$ (Fig. 2(a), inset and Fig. S1(a))}.
\textcolor{black}{Once the multipolar Kondo effect kicks in below $\Delta_{\rm CEF}\sim60$ K, the MR develops a crossover from the low-field $B^2$ behavior to a quasi-linear field dependence (Fig. 2(a))}; the crossover shifts to a lower field on cooling.
Below $T_Q\sim 2$ K, the window of $B^2$ behavior completely vanishes, and unsaturated quasi-linear MR persists up to 16 T.
Remarkably, the magnitude of MR undergoes three orders of magnitude enhancement across the FQ transition, reaching $\sim 10^3\%$ at 0.1 K (Fig. 2(a)); this value falls in the typical range $10^3-10^8\%$ of extremely large magnetoresistance (XMR)\cite{niu2021materials}. \textcolor{black}{The MR observed for $B\parallel[001]$ is nearly identical to that for $B\parallel[111]$ for \textcolor{black}{$T\gtrsim 10$ K} (Fig. 2(c)). Below $T_Q$, XMR on the order of $\sim 10^3\%$ also emerges for $B\parallel[001]$, but with clear anisotropy compared with $B\parallel[111]$ (Fig. 2(a), main panel), which will be discussed in detail later.}

Though XMR is extensively explored in topological and two-dimensional materials, it has not yet been reported in a pure orbital ordered phase. Thus, identifying the mechanism behind the observed XMR in PrTi$_{2}$Al$_{20}$ may help widen the scope of XMR studies. To explore the mechanism behind the observed XMR, we make a comparison with the magnetotransport behavior in non-4$f$ analog LaTi$_{2}$Al$_{20}$.
As shown in Fig. 2(b) and Fig. S1(b), \textcolor{black}{for both $B\parallel[111]$ and $B\parallel[001]$, the MR of LaTi$_{2}$Al$_{20}$ follows $B^2$ behavior up to 9 T for $T\gtrsim$ 200 K and develops a crossover from $B^2$ to nearly $B$-linear dependence on cooling; unsaturated XMR on the order of $10^3\%$ takes place in the FL state below 20 K (see Fig. 1, inset for the FL fit).}
Owning to the lack of 4$f$ multipolar moments in LaTi$_{2}$Al$_{20}$, this close resemblance suggests that the XMR arises from some common characteristics of the two materials that are insensitive to the presence of multipolar moments and their long-range FQ order at the low-$T$ limit. \textcolor{black}{Moreover, the $c$-$f$ hybridization tends to decline sharply in the FQ ordered state, evident from the previous specific heat measurements showing that the residual entropy associated with the quadrupolar Kondo effect is released in the FQ ordered state, as well as the gapped behavior (i.e., exponential decay) of $\rho (T)$ (Fig.1) and specific heat below the FQ transition temperature $T_Q$ \cite{Sakai2011}. Thus, it appears unlikely that the interaction between multipolar moments and conduction electrons plays a major role in producing the XMR in the FQ ordered state of PrTi$_{2}$Al$_{20}$.}

Electron-hole compensation and \textcolor{black}{ultrahigh carrier mobility (in the range of $10^4$-$10^6$ cm$^2$/Vs) are the two most common mechanisms that generate XMR\cite{niu2021materials}.
These are often realized in semimetals featuring small Fermi pockets and low density of states (DOS) near the Fermi level\cite{liang2015ultrahigh,shekhar2015extremely,yu2017magnetoresistance,mondal2020extremely,lv2018mobility,kumar2017large}.
However, both PrTi$_{2}$Al$_{20}$ and LaTi$_{2}$Al$_{20}$ are metallic systems with sizable FS sheets and DOS near the Fermi level\cite{dHvA}, unlikely to reach the ultrahigh mobility as in the semimetal cases. Moreover, carrier compensation or high mobility alone typically yields unsaturated, quadratic $\mathrm{MR}\sim B^2$, inconsistent with the observed quasi-linear field dependence shown in Fig. 2. Exceptions may occur for topological Dirac and Weyl semimetals, in which the linear band dispersion can lead to unsaturated linear-in-$B$ MR\cite{liang2015ultrahigh,shekhar2015extremely}, while this situation does not apply to the band structures of PrTi$_{2}$Al$_{20}$ or LaTi$_{2}$Al$_{20}$. 
Thus, carrier compensation or high mobility is insufficient to explain the nonsaturating linear XMR observed here; another factor is at play for its generation.} 

\textcolor{black}{Given that PrTi$_{2}$Al$_{20}$ has no spin degrees of freedom in its CEF ground state and shows no trace of charge order at low-$T$s\cite{Taniguchi2016,Taniguchi2019,koseki2011ultrasonic}, we can also rule out spin fluctuations and charge density waves as possible mechanisms for the observed XMR. We then turn to the FS properties given that the previous dHvA study suggests similar FS geometry for PrTi$_{2}$Al$_{20}$ and LaTi$_{2}$Al$_{20}$. In particular, we consider the possibility of enhanced MR due to open-orbit FS topology\cite{Wu2020,Hasegawa1991,umehara1991}.}
The previously reported FS of LaTi$_{2}$Al$_{20}$ comprises a large jungle-gym-like electron sheet (96th band), with cubic symmetry and \textcolor{black}{"necks"} along the eight symmetrically equivalent $\langle111\rangle$ axes \cite{dHvA}, bearing similarity with the well-studied FS of copper\cite{pippard1989magnetoresistance,zhang2019magnetoresistance}.
\textcolor{black}{Such "necks" in the FS can induce open orbits for the FS cross section in a certain range of magnetic field, leading to complex field angle dependence of the XMR. 
In the copper case, when $B\parallel\langle001\rangle$, the MR value is at a minimum with a convex upward curvature. As $B$ rotates from $\langle001\rangle$ to $\langle110\rangle$, the MR can exhibit unsaturated, quasi-linear field dependence for some intermediate angles (such as $B\parallel\langle111\rangle$) before reaching a maximum with quadratic field dependence \cite{pippard1989magnetoresistance,zhang2019magnetoresistance}.
Thus, the open-orbit FS topology well accounts for the quasi-linear XMR and its anisotropy between $B\parallel$ [111] and $B\parallel$ [001] at the low-$T$ limit.}
The similar MR behavior observed in PrTi$_{2}$Al$_{20}$ and LaTi$_{2}$Al$_{20}$ at the low-$T$ limit suggests that such open-orbit FS topology persists even with $c$-$f$ hybridization and the long-range FQ order, serving as a key driver of the nonsaturating, linear XMR in PrTi$_{2}$Al$_{20}$.

\textcolor{black}{We note that quantum oscillations emerge in MR and Hall effect below $\sim 1$ K in PrTi$_{2}$Al$_{20}$ \textcolor{black}{(see details in the Supplementary Materials and Fig. S7)}. The observable quantum oscillations indicate that the system fulfills the high-field limit in the FQ ordered state, namely, $\omega_c\tau\gtrsim2\pi$, where $\omega_c$ is the cyclotron frequency, and $\tau$ is the scattering time. Under this condition, the carriers can traverse a complete cyclotron orbit before being scattered, allowing the open-orbit FS geometry to be reflected in the magnetotransport, consistent with the presence of open-orbit-induced linear XMR within the FQ state. Moreover, the crossover in MR from $B^2$ to $B$-linear behavior on cooling leads to a violation of ordinary and extended Kohler's rules (Supplementary Materials and Fig. S2). Violation of the  Kohler's rule is typically associated with changes in the anisotropy pattern of the scattering time $\tau (\mathbf{k})$ on the FS, as reported in cuprates and iron-based superconductors\cite{ginsberg1998physical,kasahara2010evolution}. This scenario is again in line with more carriers moving along the open-orbit trajectories on the FS as inelastic and orbital-fluctuation-induced scatterings cease in the low-$T$ FQ state.}

Interestingly, the MR of PrTi$_{2}$Al$_{20}$ is nearly isotropic at high $T$s but becomes strongly anisotropic on approaching the FQ order \textcolor{black}{(Figs. 2(c) and (d))}: For $T_Q\lesssim T < 10$ K, the MR increases monotonically under $B\parallel [111]$, whereas forming a broad maximum and then declines at higher fields under $B\parallel [001]$. 
Within the FQ order, the MR curve displays a kink around $B = 2$ T $\parallel$ [001] \textcolor{black}{(Fig. 2(a))}, whereas this anomaly is absent for $B\parallel$ [111]; \textcolor{black}{the difference in magnitude reaches nearly two-fold at 16 T for the two field directions.
The low-field MR anomaly overlaps with a field-induced transition previously detected by NMR and magnetization around $B = 2$ T $\parallel$ [001], whereas such transition is absent for $B\parallel [111]$ \cite{Taniguchi2019,kittaka2020field}}. \textcolor{black}{We note that the MR exhibits weak RRR dependence for high-quality single crystals (Fig. S4), and thus the observed anisotropy is intrinsic rather than caused by the difference in sample quality.}
\textcolor{black}{The low-temperature MR anisotropy observed in PrTi$_{2}$Al$_{20}$ (Fig. 2(a) and (d)) is more pronounced than the anisotropy observed in LaTi$_{2}$Al$_{20}$ (Fig. 2(b) and Fig. S1(b)), suggesting that the 4$f$ multipolar moments, along with their remaining interaction with the conduction electron sea in the ordered state could offer an additional mechanism for the magnetotransport anisotropy in PrTi$_{2}$Al$_{20}$ other than the open-orbit FS topology.}

\begin{figure}[t]
\begin{center}
\includegraphics[keepaspectratio, width=8.5cm]{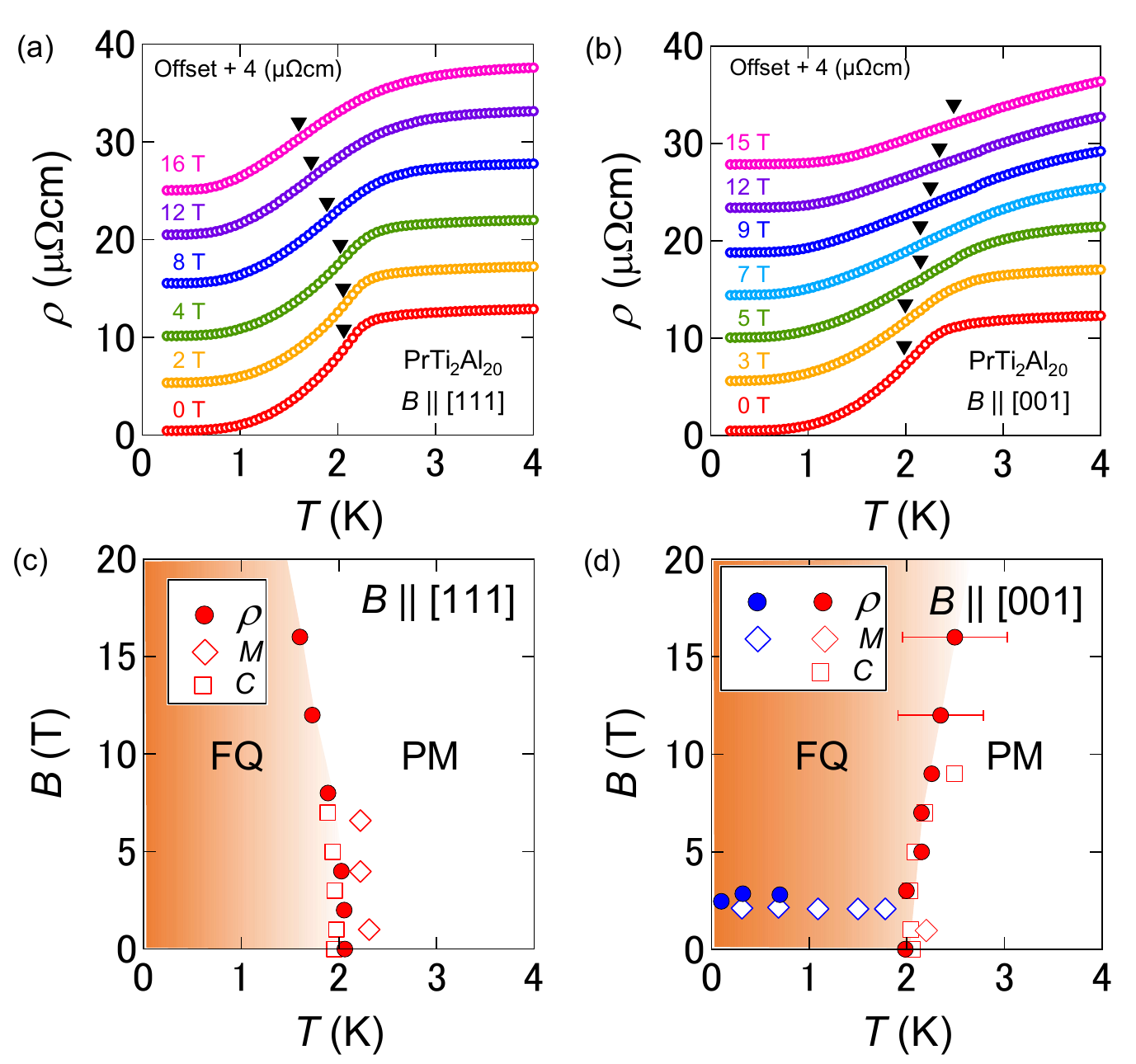}
\caption{\textcolor{black}{
(a) and (b) Temperature $T$ dependence of the resistivity $\rho (T)$ in PrTi$_{2}$Al$_{20}$ measured below 4 K under $B \parallel$ [111] (a) and $B \parallel$ [001] (b).
The curves are vertically shifted by 4 $\mu\Omega$cm/K for each finite field value.
Downward black triangles mark the FQ transition determined from the peaks of the resistivity derivative $d\rho/dT$ (see Fig. S3 in the Supplemental Materials.)
(c) and (d) The temperature-field ($T$-$B$) phase diagram of PrTi$_2$Al$_{20}$ obtained in $B\parallel$ [111] (c) and $B\parallel$ [001] (d). The red solid circles represent the FQ-PM phase boundary derived from the temperature dependence of $\rho$ (i.e., the black triangles in (a) and (b)). The blue solid circles represent the field-induced transition around 2 T obtained from the MR measurements;} the open symbols are based on previously reported magnetization \textcolor{black}{($M$: open diamonds)} and specific heat measurements \textcolor{black}{($C$: open squares)} \cite{Sakai2011, Taniguchi2019}}
\label{fig_Hall_analysis}
\end{center}
\end{figure}

To further investigate the anisotropic low-$T$ transport in PrTi$_{2}$Al$_{20}$, we outline the highly anisotropic temperature-field ($T$-$B$) phase diagrams under $B\parallel [111]$ and $B\parallel [001]$ based on isofield resistivity measurements \textcolor{black}{(Figs. 3(a), (b) and Figs. S3).}
\textcolor{black}{The phase diagrams based on transport data \textcolor{black}{shown in Figs. 3(c) and (d)} agrees with those obtained by previous NMR and thermodynamic measurements\cite{Sakai2011,Taniguchi2019,kittaka2020field}. \textcolor{black}{For $B\parallel [111]$,} the FQ transition remains well-defined up to $\sim 10$ T, with $T_Q$ decreasing as $B$ rises beyond 5 T \textcolor{black}{(Fig. 3(a), Fig. 3(c) and Fig. S3(a), (c))}.
In contrast, the long-range FQ order is "soft" for $B\parallel [001]$. A small field of $B \sim 2$ T $\parallel [001]$ turns the second-order FQ transition into a smooth crossover, evidenced by the drastic broadening of the resistivity anomaly at $T_Q$ \textcolor{black}{(Fig. 3(b) and Fig. S3(b), (d))}; the FQ-paramagnet (PM) phase boundary bends toward higher $T$ with increasing $B$ \textcolor{black}{(Fig. 3(d))}.
The observed anisotropic field evolution of \textcolor{black}{the FQ-PM transition boundary} results from the strongly anisotropic response of the FQ order to [111] and [001] magnetic fields, which can be explained by the competition between the Zeeman effect and the field-induced quadrupolar-quadrupolar interaction\cite{Taniguchi2019}.
The possibility for a metastable domain structure is also discussed theoretically \cite{lee2018landau}.
The quadrupolar short-range fluctuations associated with the crossover under $B \parallel$ [001] can alter the carrier scattering and thereby the magnetotransport near $T_Q\sim 2$ K, leading to the sharply different behavior of MR compared to that in $B \parallel$ [111].}

Below $T_Q\sim 2$ K, the MR anomaly near 2 T for $B \parallel$ [001] \textcolor{black}{(Fig. 2(a))} is likely associated with a change in the quadrupolar ordering structure\cite{Taniguchi2019,kittaka2020field, freyer2020thermal}. Earlier comparison of the NMR and specific heat results with theoretical analysis suggests that the low-field transition might be driven by FQ order parameter switching accompanied by a change in the $c$-$f$ hybridization, which may cause FS reconstruction\cite{Taniguchi2019, kittaka2020field}.
Then the difference in quadrupolar ordering structure between [111] and [001] direction can induce the transport anisotropy via modifying the anisotropy pattern of the scattering rate on the FS.

The Hall resistivity may offer more insight into the interplay of multipolar order with FS and transport properties. \textcolor{black}{PrTi$_{2}$Al$_{20}$ is ideal for investigating the Hall effect in a multipolar heavy fermion system, which has yet to be explored.}
\textcolor{black}{Figure 4(a)} shows the magnetic field $B$ dependence of the Hall resistivity $\rho_{\rm H}$ for PrTi$_{2}$Al$_{20}$ at various temperatures and the $B$ dependence of the Hall coefficient $R_{\rm H}\equiv \rho_{\rm H}/B$ of PrTi$_{2}$Al$_{20}$ obtained at selected temperatures shown in \textcolor{black}{Fig. 4(c)}.
$R_{\rm H}$ shows clear $B$ dependence, reflecting nonlinear $\rho_{\rm H}$ as a function of $B$. Such nonlinearity indicates the multiple-band signature, as discussed below. \textcolor{black}{The high-field anomaly at $\sim$ 11 T (Figs. 4(a) and (c)) is likely related to a field-induced rearrangement of multipolar moments within the FQ ordered phase, similar to the transition reported for the sister compound PrV$_{2}$Al$_{20}$ under $B\sim 11$ T $\parallel [001]$ \cite{shimura2013evidence,shimura2019giant}. Further experiments on detecting the FQ order parameter are necessary to confirm this scenario.}

Next, we explore the temperature dependence of the initial Hall coefficient $R_{\rm H}^0$ for PrTi$_{2}$Al$_{20}$ \textcolor{black}{(Fig. 4(e))}, where $R_{\rm H}^0$ is defined as the slope of the Hall resistivity versus field isothermals at the zero-field limit.
$R_{\rm H}^0$ is positive in the entire measured $T$ range, indicating hole-type majority charge carriers.
With decreasing $T$, $R_{\rm H}^0$ exhibits a mild increase below $\Delta_{\mathrm{CEF}} \sim 60$ K, then forming a plateau in the FL state \textcolor{black}{(7 K $\lesssim T \lesssim 17$ K)}; a more pronounced upturn of $R_{\rm H}^0$ occurs near $T_Q\sim 2$ K, followed by saturation in the ordered state. 

Such temperature dependence of $R^0_{\rm H}$ is qualitatively distinct from the behavior typically seen in magnetic heavy fermion metals\cite{Fert1987}.
\textcolor{black}{
The Hall coefficient consists of the normal Hall component $R_{H}^{\rm N}$ and the anomalous Hall component $R_{\rm H}^{\rm A}$.}
As shown by the dashed lines in \textcolor{black}{Fig. 4(e)}, the initial Hall coefficient $R^0_H$ of various classes of magnetic heavy fermion compounds displays a common feature\cite{Fert1987}: a broad maximum near the coherence temperature $T_{\rm coh}$, marking a crossover from the high-$T$ regime dominated by the skew-scattering-induced anomalous Hall effect $R_{\rm H}^{\rm A}$ to the low-$T$ coherence regime where 4$f$ moments enters the Fermi volume, forming a heavy FL. In the high-$T$ regime, the $R_{\rm H}^{\rm A}$ is well-scaled by the magnetic susceptibility, such that $R_{\rm H}^{\rm A}\propto\chi$ or $R_{\rm H}^{\rm A}\propto\chi\rho$, where $\rho$ is the longitudinal resistivity. In contrast, the coherence peak is absent in $R^0_H$ of PrTi$_{2}$Al$_{20}$, and $R^0_H$ cannot be scaled by either $\chi$ or $\chi\rho$ (\textcolor{black}{Figs. 5(a), (b) and Fig. S5})\cite{Fert1987}. \textcolor{black}{Thus, the anomalous Hall component associated with the magnetization is negligibly small at least below 100 K in PrTi$_{2}$Al$_{20}$, as expected from the lack of dipolar degrees of freedom in its CEF ground-state doublet and the sizeable gap $\Delta_\mathrm{CEF}\sim 60$ K between the ground-state doublet and the first-excited magnetic triplet.} Moreover, the overall field dependence and magnitude of $\rho_{\rm {H}}$ and the Hall coefficient $R_{\rm H} \equiv \rho_{\rm H}/B$ in the FQ ordered state of PrTi$_{2}$Al$_{20}$ are similar to that of LaTi$_{2}$Al$_{20}$ \textcolor{black}{(Fig. 5(a))}, indicating that the anomalous Hall contribution from the high-rank multipolar moments has minor influence on the low-$T$ Hall effect. \textcolor{black}{Altogether, our findings indicate that $R{\rm_H}$ of PrTi$_{2}$Al$_{20}$ is governed by the normal Hall contribution $R_{\rm H}^{\rm N}$, same as in LaTi$_{2}$Al$_{20}$, which allows comparison of the Hall coefficient behavior in the two systems.}

\begin{figure}[t]
\centering 
\includegraphics[keepaspectratio, width=9cm]{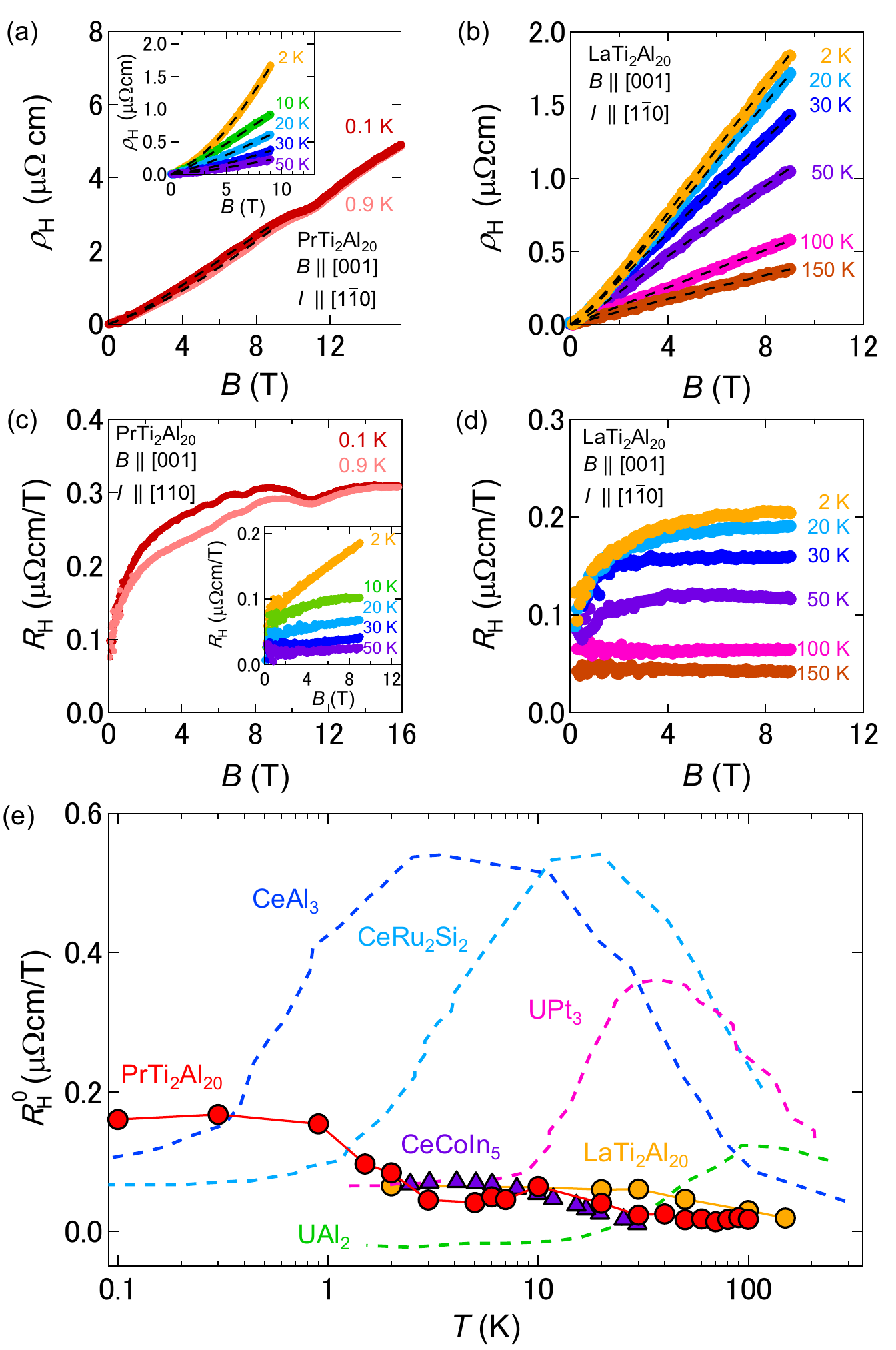}
\caption{(a) Magnetic field $B$ dependence of the Hall resistivity $\rho_{\rm H}$ and (b) the Hall coefficient $R_{\rm H}$ of PrTi$_2$Al$_{20}$ obtained at representative temperatures in $B\parallel [001]$.
The main panel shows $\rho_{\rm H}$ obtained within the FQ ordered phase at 0.1 K and 0.9 K, and the inset shows $\rho_\mathrm{H}$ measured above the FQ transition.
\textcolor{black}{The black dashed lines represent the two-band fits performed for 0 T to 9 T, avoiding the high-field anomaly centered at $\sim 11$ T.}
(c) Magnetic field $B$ dependence of $\rho_\mathrm{H}$ and (d) $R_\mathrm{H}$ of LaTi$_2$Al$_{20}$ measured under $B\parallel [001]$ in the temperature range from 2 K to 150 K.
(e) The initial Hall coefficient $R_{\rm H}^0$ of PrTi$_2$Al$_{20}$ (red solid symbols) and LaTi$_2$Al$_{20}$ (orange solid symbols) obtained from the slope of the isothermal $\rho_{\rm H}$ vs.$B$ curves in the zero-field limit.
Dashed lines are $R_{\rm H}^0$ for representative Ce- and U-based heavy fermion compounds\cite{Fert1987}.
}
\label{fig_Hall} 
\end{figure}

The normal Hall effect yields information about carrier density and mobility.
\textcolor{black}{
In the high-field limit $\omega_c\tau \gtrsim 2\pi$, the Hall coefficient $R_{\rm H}$ reaches a field-independent value, $R_{\rm H}(\infty) =\frac{1}{e(n_h-n_e)}$, that provides a measure of the net carrier density enclosed by the FS.
}
For both $B\parallel [001]$ and $[111]$ , the $\rho_{\rm H}$ of PrTi$_{2}$Al$_{20}$ becomes nonlinear in $B$ for $T_Q < T < 50$ K, and its magnitude is strongly $T$-dependent \textcolor{black}{(Fig. 4(a), inset and Fig. S6(a), inset)}.
Such nonlinearity is more pronounced than that observed in LaTi$_{2}$Al$_{20}$ \textcolor{black}{(Fig. 4(b) and Fig. S6(c))}, suggesting carrier mobility misbalances on the electron and hole FS sheets due to the interplay between 4$f$ quadrupolar moments and conduction electrons. Moreover, $R_{\rm H}$ of PrTi$_{2}$Al$_{20}$ does not fully saturate up to 9 T \textcolor{black}{(Fig. 4(c), inset and Fig. S6(b), inset)}, in contrast with $R_{\rm H}$ in LaTi$_{2}$Al$_{20}$ that levels off at $\sim 3$ T (\textcolor{black}{Fig. 4(d) and Fig. S6(d)}); namely, a higher-$B$ is necessary for PrTi$_{2}$Al$_{20}$ to reach the high-field limit. These facts suggest that the transport properties above $T_Q$ are strongly affected by substantial quadrupolar fluctuations as an additional scattering mechanism. 
This point is corroborated by the above-mentioned MR crossover in PrTi$_{2}$Al$_{20}$ from low-field $B^2$-dependence to quasi-$B$-linear dependence without saturation in the same $T$ range (Fig. 2(a), inset and Fig. S1(a)).

With suppressed quadrupolar fluctuations below $T_Q$, the isothermal field-sweeps of $\rho_{\rm H} (B)$ become nearly $T$-independent but still exhibits nonlinearity (\textcolor{black}{Fig. 4(a), main panel}).
\textcolor{black}{
For $B\parallel [001]$, the Hall coefficient $R_{\rm H}$ (Fig. 4(c), main panel and Fig. 5(a)) displays a smooth crossover from the low-field increase to a field-independent regime above $\sim 8$ T; the high-field $R_{\rm H}$ value is larger than that of LaTi$_{2}$Al$_{20}$, indicating a lower net carrier density in the FQ ordered state.
This feature indicates that the 4$f$ moments only weakly contribute to the Fermi volume below $T_Q$.
Unlike the magnetic susceptibility shown in Fig. 5(b), $R_{\rm H}$ does not exhibit any sharp anomaly at $B\sim 2$ T $\parallel [001]$; therefore, the field-induced change in quadrupolar ordering structure does not seem to trigger noticeable FS reconstruction, possibly owing to the strongly suppressed $c$-$f$ hybridization below $T_Q$\cite{Sakai2011}. 
}

\textcolor{black}{The multiband effect is a typical mechanism for the non-linear $B$ dependence of the normal Hall effect. We therefore estimate the carrier densities and mobilities by fitting $\rho_{\rm H}$ to the widely-used two-band model that simplifies multi-band scenario into one electron-type band and one hole-type band\cite{pippard1989magnetoresistance, Chambers1952}, such that}
\begin{equation}
\rho_{\rm H}(B)=\frac{B}{e}\frac{(n_{h}\mu^{2}_{h}-n_{e}\mu^{2}_{e})+(n_{h}-n_{e})\mu^{2}_{e}\mu^{2}_{h}B^2}{(n_{h}\mu_{h}+n_{e}\mu_{e})^2+(n_{h}-n_{e})^2\mu^{2}_{e}\mu^{2}_{h}B^2},
\end{equation}
\textcolor{black}{where $n_{e,h}$ are the electron-type and hole-type carrier densities, and $\mu_{e,h}$ are their respective mobilities (see details in the Supplementary Materials).
Though the two-band fitting cannot fully capture the complex multiband FS of PrTi$_{2}$Al$_{20}$ and LaTi$_{2}$Al$_{20}$, it reasonably estimates the carrier properties (see Supplementary Table 1) \cite{dHvA}.} \textcolor{black}{We note that the field range of the fits is limited to 0 T $\leq B\leq 9$ T to avoid inclusion of the high-field anomaly located around $\sim 11$ T.}
The electron and hole densities and mobilities obtained from the two-band fit are shown in Figs. 5(c) and (d), compared with the results estimated from the initial Hall coefficient $R^0_H$ (i.e., the one-band model). 
The increasing deviation between the two-band and one-band fitting results reflects the nonlinear field dependence of $\rho_{\rm H}$ at low $T$s, originating from the multiband effect.

For both PrTi$_{2}$Al$_{20}$ and LaTi$_{2}$Al$_{20}$, the hole band dominates the electrical transport process at all measured $T$s. The electron-like contribution $n_e$ becomes appreciable near $T_Q$ in PrTi$_2$Al$_{20}$ (Fig. 5(c)); the electron and hole mobilities, $\mu_e$ and $\mu_h$, are both dramatically enhanced on cooling below $T_Q$, then leveling off in the FQ ordered state (Fig. 5(d)).
\textcolor{black}{These results support the scenario that the suppression of quadrupolar-fluctuation-induced scatterings below $T_Q$ leads to sufficiently high carrier mobilities, which are necessary for observing the linear XMR induced by open-orbit trajectories on the large electron-like FS sheet.
The qualitative behavior and low-$T$ values of carrier densities and mobilities in LaTi$_{2}$Al$_{20}$ resemble the behavior in PrTi$_2$Al$_{20}$.
A key difference is that the enhancement of electron-band contribution appears at much higher temperatures ($T \lesssim 70$ K) in LaTi$_{2}$Al$_{20}$ (Fig. 5(c) and (d)), likely due to the lack of quadrupolar fluctuations.
This difference naturally explains why the onset of XMR occurs at a lower temperature in PrTi$_{2}$Al$_{20}$ than in LaTi$_{2}$Al$_{20}$.}
Moreover, the electron and hole mobilities overlap in  LaTi$_{2}$Al$_{20}$ but differ ($\mu_e < \mu_h$) at low-$T$ regime in PrTi$_{2}$Al$_{20}$, which explains the stronger nonlinearity of $\rho_{\rm H} (B)$ in PrTi$_{2}$Al$_{20}$. The imbalanced $\mu_e$ and $\mu_h$ in  PrTi$_{2}$Al$_{20}$ might result from the mildly enhanced $m^*$ on the electron FS, the anisotropic scattering time $\tau$ induced by the quadrupolar order parameter or both.

Similar to the MR, the Hall resistivity $\rho_{\rm H} (B)$ in the FQ ordered phase of  PrTi$_{2}$Al$_{20}$ is strongly anisotropic in [001] and [111] magnetic fields: For $B\parallel [111]$, the $\rho_{\rm H} (B)$ shows concave curvature above $\sim 4$ T, and its magnitude at 0.1 K is only about \textcolor{black}{one-third} of that for $B\parallel [001]$ \textcolor{black}{(Figs. S6(a) and (b))}.
The simple two-band scenario cannot describe the field dependence of $\rho_{\rm H} (B)$ under $B\parallel [111]$.
Again, such a pronounced anisotropy is absent in LaTi$_{2}$Al$_{20}$ \textcolor{black}{(Figs. 4(b), (d), Figs. S6(c) and (d))},
implying that the anisotropic transport in  PrTi$_{2}$Al$_{20}$ originates from the field-induced change of FQ order structure under $B\parallel [001]$.

\begin{figure}[t]
\begin{center}
\includegraphics[keepaspectratio, width=8.5cm]{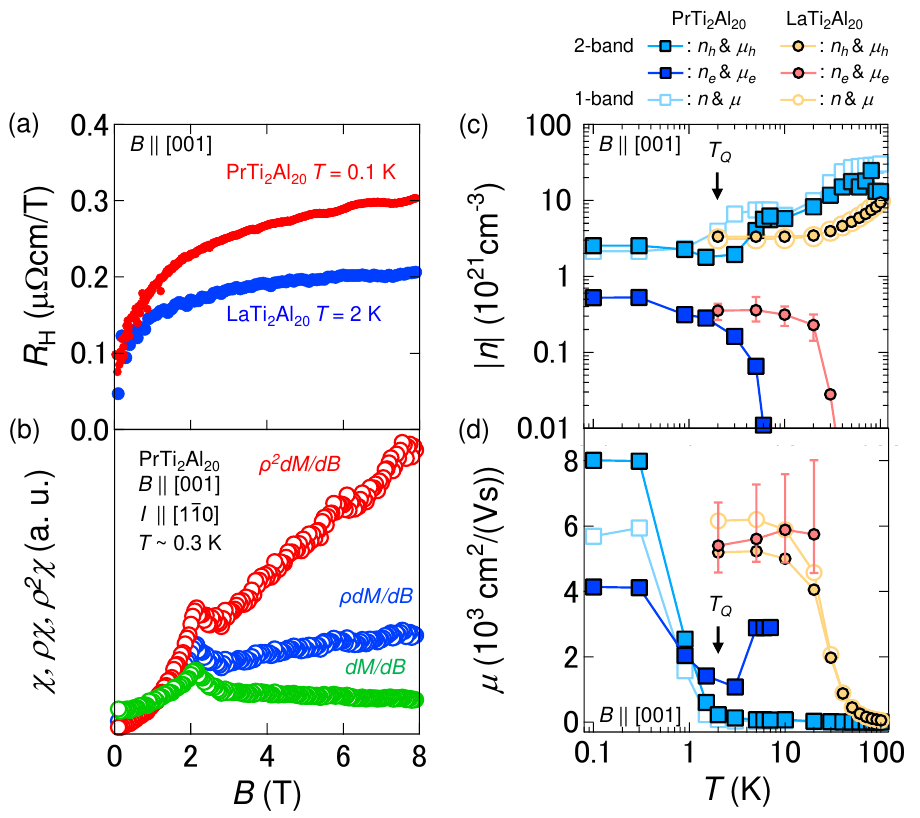}
\caption{(a) Low-$T$ Hall coefficient $R_{\mathrm{H}}$ vs $B$ of PrTi${_2}$Al$_{20}$ and LaTi$_{2}$Al$_{20}$ compared with (b) the $B$ scaling of the magnetic susceptibility $\chi$ and the quantities $\chi\rho$ and $\chi\rho^2$ of PrTi$_{2}$Al$_{20}$ measured at 0.3 K.
(c) The evolution of the charge carrier density $|n|$ and (d) mobility $\mu$ as a function of temperature $T$ estimated based on \textcolor{black}{the two-band fits with fitting range 0 T $\leq B\leq 9$ T} (solid symbols) \textcolor{black}{in LaTi$_{2}$Al$_{20}$ ($n_e$ and $\mu_e$: orange, $n_h$ and $\mu_h$: yellow) and PrTi$_{2}$Al$_{20}$ ($n_e$ and $\mu_e$: dark blue, $n_h$ and $\mu_h$: light blue)} and from the initial Hall coefficient $R_{\rm H}^0$ (i.e. one-band fitting) \textcolor{black}{in LaTi$_{2}$Al$_{20}$ (open yellow circles) and PrTi$_{2}$Al$_{20}$ (open light blue circles). The one-band fitting yields dominated hole-type charge carriers, consistent with the two-band fitting results.}}
\label{fig_Hall_5}
\end{center}
\end{figure}

\textcolor{black}{To summarize, our comprehensive study of magnetotransport of PrTi$_2$Al$_{20}$ reveals extremely large magnetoresistance (XMR)  reaching $\sim 10^{3}\%$ in its FQ ordered state. Based on comparison with the non-4$f$ analog, LaTi$_{2}$Al$_{20}$, we attribute this XMR to the open-orbit topology of the Fermi surface. The $B$-dependence of the Hall resistivity $\rho_{\rm H}$ displays stronger nonlinearity than that of LaTi$_{2}$Al$_{20}$ on approaching the FQ state.
Analysis using the two-band model indicates that the contribution from the electron-type FS sheet featuring open orbits becomes effective upon suppression of the quadrupolar-fluctuation scattering, further supporting the FS topology's key role in generating the XMR. Both the MR and $\rho_{\rm H}$ become highly anisotropic under $B\parallel [111]$ and $B\parallel [001]$ in the FQ state, following the distinct response of the FQ order parameter under the two field orientations.}
\textcolor{black}{Our study indicates that multipolar ordered state without involving spin degrees of freedom can realize large magnetoresistance.
These findings provide essential insights that may help identifying universal mechanisms behind large magnetotransport phenomena and thereby widen their applications.}

This work is partially supported by CREST (JPMJCR18T3), Japan Science and Technology Agency, by Grants-in-Aids for Scientific Research on Innovative Areas (15H05882 and 15H05883) from the Ministry of Education, Culture, Sports, Science and Technology of Japan, and by Grants-in-Aid for Scientific Research (19H00650) from the Japanese Society for the Promotion of Science (JSPS).
This work is partially supported by JSPS KAKENHI Grant Number 20K03829.
T.I. is supported by JST SPRING, Grant Number JPMJSP2108.
M.F. acknowledges support from the Japan Society for the Promotion of Science Postdoctoral Fellowship for Research in Japan (Standard).

\bibliographystyle{PRResearch}

\end{document}